\documentstyle[12pt]{article}

\advance \topmargin by -\headheight
\advance \topmargin by -\headsep
     
\evensidemargin \oddsidemargin
 \date{}    
\begin{document}
\newcommand{\sect}[1]{\setcounter{equation}{0}\section{#1}}
\renewcommand{\theequation}{\thesection.\arabic{equation}}

\topmargin -.6in
\def\nonu{\nonumber}
\def\rf#1{(\ref{eq:#1})}
\def\lab#1{\label{eq:#1}} 
\def\br{\begin{eqnarray}}
\def\er{\end{eqnarray}}
\def\be{\begin{equation}}
\def\ee{\end{equation}}
\def\0{\nonumber}

\def\lb{\lbrack}
\def\rb{\rbrack}
\def\({\left(}
\def\){\right)}
\def\v{\vert}
\def\bv{\bigm\vert}
\def\lskip{\vskip\baselineskip\vskip-\parskip\noindent}
\relax
\newcommand{\nit}{\noindent}
\newcommand{\ct}[1]{\cite{#1}}
\newcommand{\bi}[1]{\bibitem{#1}}
\def\a{\alpha}
\def\b{\beta}
\def\ca{{\cal A}}
\def\cm{{\cal M}}
\def\cn{{\cal N}}
\def\cf{{\cal F}}
\def\d{\delta} 
\def\D{\Delta}
\def\eps{\epsilon}
\def\g{\gamma}
\def\G{\Gamma}
\def\grad{\nabla}
\def\h{ {1\over 2}  }
\def\hc{\hat{c}}
\def\hd{\hat{d}}
\def\hg{\hat{g}}
\def\hp{ {+{1\over 2}}  }
\def\hm{ {-{1\over 2}}  }
\def\k{\kappa}
\def\l{\lambda}
\def\L{\Lambda}
\def\lg{\langle}
\def\m{\mu}
\def\n{\nu}
\def\o{\over}
\def\om{\omega}
\def\O{\Omega}
\def\p{\phi}
\def\pa{\partial}
\def\pr{\prime}
\def\ra{\rightarrow}
\def\rh{\rho}
\def\rg{\rangle}
\def\s{\sigma}
\def\t{\tau}
\def\th{\theta}
\def\ti{\tilde}
\def\wti{\widetilde}
\def\inte{\int dx }
\def\xb{\bar{x}}
\def\yb{\bar{y}}

\def\tr{\mathop{\rm tr}}
\def\Tr{\mathop{\rm Tr}}
\def\partder#1#2{{\partial #1\over\partial #2}}
\def\ds{{\cal D}_s}
\def\wtwo{{\wti W}_2}
\def\lie{{\cal G}}
\def\alie{{\widehat \lie}}
\def\dlie{{\cal G}^{\ast}}
\def\elie{{\widetilde \lie}}
\def\edlie{{\elie}^{\ast}}
\def\hlie{{\cal H}}
\def\wlie{{\widetilde \lie}}

\def\rlx{\relax\leavevmode}
\def\inbar{\vrule height1.5ex width.4pt depth0pt}
\def\IZ{\rlx\hbox{\sf Z\kern-.4em Z}}
\def\IR{\rlx\hbox{\rm I\kern-.18em R}}
\def\IC{\rlx\hbox{\,$\inbar\kern-.3em{\rm C}$}}
\def\one{\hbox{{1}\kern-.25em\hbox{l}}}

\def\PRL#1#2#3{{\sl Phys. Rev. Lett.} {\bf#1} (#2) #3}
\def\NPB#1#2#3{{\sl Nucl. Phys.} {\bf B#1} (#2) #3}
\def\NPBFS#1#2#3#4{{\sl Nucl. Phys.} {\bf B#2} [FS#1] (#3) #4}
\def\CMP#1#2#3{{\sl Commun. Math. Phys.} {\bf #1} (#2) #3}
\def\PRD#1#2#3{{\sl Phys. Rev.} {\bf D#1} (#2) #3}
\def\PLA#1#2#3{{\sl Phys. Lett.} {\bf #1A} (#2) #3}
\def\PLB#1#2#3{{\sl Phys. Lett.} {\bf #1B} (#2) #3}
\def\JMP#1#2#3{{\sl J. Math. Phys.} {\bf #1} (#2) #3}
\def\PTP#1#2#3{{\sl Prog. Theor. Phys.} {\bf #1} (#2) #3}
\def\SPTP#1#2#3{{\sl Suppl. Prog. Theor. Phys.} {\bf #1} (#2) #3}
\def\AoP#1#2#3{{\sl Ann. of Phys.} {\bf #1} (#2) #3}
\def\PNAS#1#2#3{{\sl Proc. Natl. Acad. Sci. USA} {\bf #1} (#2) #3}
\def\RMP#1#2#3{{\sl Rev. Mod. Phys.} {\bf #1} (#2) #3}
\def\PR#1#2#3{{\sl Phys. Reports} {\bf #1} (#2) #3}
\def\AoM#1#2#3{{\sl Ann. of Math.} {\bf #1} (#2) #3}
\def\UMN#1#2#3{{\sl Usp. Mat. Nauk} {\bf #1} (#2) #3}
\def\FAP#1#2#3{{\sl Funkt. Anal. Prilozheniya} {\bf #1} (#2) #3}
\def\FAaIA#1#2#3{{\sl Functional Analysis and Its Application} {\bf #1} (#2)
#3}
\def\BAMS#1#2#3{{\sl Bull. Am. Math. Soc.} {\bf #1} (#2) #3}
\def\TAMS#1#2#3{{\sl Trans. Am. Math. Soc.} {\bf #1} (#2) #3}
\def\InvM#1#2#3{{\sl Invent. Math.} {\bf #1} (#2) #3}
\def\LMP#1#2#3{{\sl Letters in Math. Phys.} {\bf #1} (#2) #3}
\def\IJMPA#1#2#3{{\sl Int. J. Mod. Phys.} {\bf A#1} (#2) #3}
\def\AdM#1#2#3{{\sl Advances in Math.} {\bf #1} (#2) #3}
\def\RMaP#1#2#3{{\sl Reports on Math. Phys.} {\bf #1} (#2) #3}
\def\IJM#1#2#3{{\sl Ill. J. Math.} {\bf #1} (#2) #3}
\def\APP#1#2#3{{\sl Acta Phys. Polon.} {\bf #1} (#2) #3}
\def\TMP#1#2#3{{\sl Theor. Mat. Phys.} {\bf #1} (#2) #3}
\def\JPA#1#2#3{{\sl J. Physics} {\bf A#1} (#2) #3}
\def\JSM#1#2#3{{\sl J. Soviet Math.} {\bf #1} (#2) #3}
\def\MPLA#1#2#3{{\sl Mod. Phys. Lett.} {\bf A#1} (#2) #3}
\def\JETP#1#2#3{{\sl Sov. Phys. JETP} {\bf #1} (#2) #3}
\def\JETPL#1#2#3{{\sl  Sov. Phys. JETP Lett.} {\bf #1} (#2) #3}
\def\PHSA#1#2#3{{\sl Physica} {\bf A#1} (#2) #3}
\def\PHSD#1#2#3{{\sl Physica} {\bf D#1} (#2) #3}
\def\Nonlinearity#1#2#3{{\sl Nonlinearity} {\bf #1} (#2) #3}
%supervertex.tex  \hskip 8cm  \today

\begin{center}
{\large\bf  Soliton Solutions for the Super mKdV and sinh-Gordon Hierarchy}
\end{center}
\normalsize
\vskip .4in

\begin{center}
J.F. Gomes,  L.H. Ymai and A.H. Zimerman

\par \vskip .1in \noindent
Instituto de F\'{\i}sica Te\'{o}rica-UNESP\\
Rua Pamplona 145\\
01405-900 S\~{a}o Paulo, Brazil
\par \vskip .3in

\end{center}

\begin{abstract}
 The dressing and vertex operator formalism is emploied to study  the soliton solutions of the $N=1$ 
super mKdV and sinh-Gordon models.  Explicit  two and four vertex  solutions are constructed. 
The relation between the soliton solutions of both models is verified.

\end{abstract}

\section{Introduction}

A systematic   construction of supersymmetric integrable hierarchies within 
the algebraic formalism was proposed in \cite{npb}.
An interesting feature of such approach is that it allows time evolutions  
according to both, positive and negative  grades.  In particular the first 
negative grade time evolution is always associated 
to the relativistic integrable model (with appropriated choice of space-time coordinates).    
Specific examples were given in connection with the $sl(2,1)$  
super Lie algebra yielding  the super mKdV and sinh-Gordon  models. 
The supersymmetric sinh-Gordon was  proposed in \cite{chai} by introducing a pair of 
Grassmann coordinates whilst the super mKdV, involving   a single Grassmann coordinate,
 was proposed later in \cite{mathieu}. 
 By extending Hirota's method for superfields, solutions for the super mKdV 
  was obtained \cite{liu}.   
The relation between the bosonic mKdV and sinh-Gordon models  
was observed  by a number of authors \cite{sgkdv}, \cite{laf}.
This fact was explained and further extended to other integrable models in \cite{jpa} 
(e.g. Lund-Regge and Non linear Schroedinger) and in \cite{wigner} the relation between
soliton solutions of both models were verified explicitly.  This fact was explored 
in \cite{npb} to show that the super 
mKdV and sinh-Gordon  models share  the same algebraic structure and henceforth 
belong to the same integrable hierarchy.  
This is verified explicitly here by comparing the soliton solutions of both models.

In this paper we  employ the algebraic formalism to study  the $N=1$ 
super mKdV and sinh-Gordon models by decomposing the affine $\hat {sl}(2,1)$ 
into half-integer graded subspaces following ref. \cite{npb}. 
In order to obtain a general non-trivial soliton solution from the dressing formalism and vertex operators
we employ a slightly different gradation and loop automorphism from those of ref. \cite{npb}.

This paper is organized as follows. In Sect. 2 we discuss the decomposition 
of the affine $\hat sl(2,1)$ super algebra into half integer
graded subspaces and construct the mKdV and sinh-Gordon models. 
In Sect. 3 we follow refs. \cite{olive}, \cite{laf} to derive the tau functions from  the dressing formalism and 
construct the vertex operators leading to the soliton solutions
for  those integrable models.  
In the appendices we discuss the relevant subalgebra of the affine $\hat {sl}(2,1)$ and  
the general four vertex solution.

%%%%%%%%%%%%%%%%%%%%%%%%%%%%%%%%%%%%%%%%%%%%%%%%%%%%%%%%%%%%%%%%%%%%%%%%%%%%%%%%%%%%%%%%%%%%%%

\section{The super mKdV and sinh-Gordon Hierarchy}
In this section we employ the algebraic formalism to construct  an integrable  hierarchy containing 
the mKdV and sinh-Gordon supersymmetric models.
Consider the $sl(2,1)$ super Lie algebra with generators 
\br
h_1= \a_1 \cdot H, \;\; h_2 = \a_2 \cdot H,\;\;  E_{\pm \a_1}, \;\; E_{\pm \a_2},\;\; E_{\pm (\a_1+\a_2)}
\label{0.1}
\er
where $\a_1$ and $\a_2, \a_1 + \a_2$ are bosonic and fermionic roots respectively. 
The integrable hierarchy is  defined  by choosing the grading operator 
\br
Q= 2d + {1\o 2} h_1
\label{0.2}
\er
where $d$ satisfy $[d, T_a^{(m)}] = m T_a^{(m)}, \;\; T_a^{(m)}$ denote both  $ E_\a ^{(m)}$ or $H_i^{(m)}$.
The hierarchy is further specified  by the constant grade one element, $E = E^{(1)}$ where,
\br
E^{(2n+1)} = h_1^{(n+1/2)} + 2 h_2^{(n+1/2)} - E_{\a_1}^{(n)} -  E_{-\a_1}^{(n+1)}
\label{0.3}
\er
A key ingredient to construct the desired integrable models is the judicious choice of a 
subalgebra $\hat \lie $ of the affine 
$ \hat {sl}(2,1)$. This is discussed in the appendix A.
The grading operator $Q$ and $E$ decomposes the associated affine super Kac-Moody algebra 
$\hat \lie = \oplus \lie_l = {\cal K}\oplus {\cal M}$ where $l$ is the degree of the subspace $\lie_l$ and 
${\cal K} = \{ x \in \hat {\lie }, [ x, {\cal K} ]=0\}$  
denote  the kernel of $E$ and ${\cal M}$ its complement, i.e.,
\br
{\cal K}_{Bose} = \{ K_1^{(2n+1)}, K_2^{(2n+1)} \}, \quad 
\quad {\cal K}_{Fermi} = \{ F_1^{(2n+3/2)}, F_2^{(2n+1/2)} \}, \nonu \\
{\cal M}_{Bose} = \{ M_1^{(2n+1)}, M_2^{(2n)} \}, \quad 
\quad {\cal M}_{Fermi} = \{ G_1^{(2n+1/2)}, G_2^{(2n+3/2)} \}
\label{0.4}
\er
where the generators $K_i, M_i,F_i$ and $G_i$ are constructed in terms of generators of $\hat {sl}(2,1)$ in app. A.
Define the Lax operator $L= \pa_x  + E + A_0 + A_{1/2} = \pa_x + {\cal A}_x$ where 
$ {\cal A}_x = A_0  + A_{1/2} \in {\cal M} \;\; {\rm {mod}} \;\; \hat c$, i.e., 
\br 
A_0 = uM_2^{(0)} + \eta \hat c, \quad \quad A_{1/2} = \bar \psi G_1^{(1/2)}
\label{0.5}
\er
The positive hierarchy is  given in terms of the zero curvature condition 
\br 
[\pa_x + E +A_0 + A_{1/2} \; ,\;  \pa_{t_n} + D^{(n)} + D^{(n-1/2)} + \cdots D^{(0)}] = 0
\label{0.6}
\er
where $D^{(i)} \in \lie_i$ and can be solved recursively decomposing eqn. (\ref{0.6}) grade by grade 
(see for instance ref. (\cite{npb}).
The solution is local  and the image part of the zero  and one-half grade components of (\ref{0.6})
 yields the time evolution for  the fields defined in  (\ref{0.5}).   For $n=3$ we find the equations of motion for the $N=1$ super mKdV,
 i.e.,
\br
4 \pa_{t_3} \bar \psi &=& \pa_x^3 \bar \psi - 3 u \pa_x (u \bar \psi ), \nonu \\
4 \pa_{t_3} u &=& \pa_x^3 u -6 u^2 \pa_x u + 3 \bar \psi \pa_x (u \pa_x \bar \psi )
\label{0.7}
\er
Observe that the equation of motion for the field $\eta$ is not fixed due to ambiguity in  
determining $D^{(0)} \rightarrow D^{(0)} + \rho \hat c$.  In general, for non  relativistic theories, it is necessary a more restrictive
structure given by the dressing transformations explained in the next section.
Other integrable equations are obtained for different values of $n$ in similar manner.
The negative hierarchy is obtained by 
\br 
[\pa_x + E +A_0 + A_{1/2} \; ,\;  \pa_{t_{-m}} + D^{(-m)} + D^{(-m+1/2)} + \cdots D^{(-1)} + D^{(-1/2)}] = 0
\label{0.8}
\er

Both, positive and negative hierarchies (\ref{0.6}) were shown to be derived 
from the Riemann-Hilbert problem for the homogeneous gradation \cite{rh}.
The solution for the negative hierarchy (\ref{0.8}) is, in general non-local, 
however, the simplest member of the negative hierarchy for  $m=1$ in (\ref{0.8})  has a 
closed  local solution  in terms of the zero grade group element $B \in
\lie_0$, 
\br
A_0 = -\pa_x B B^{-1}, \quad A_{1/2} = \bar \psi G_1^{(1/2)}, \quad D^{(-1/2)} = \psi B  G_2^{(-1/2)} B^{-1}, \quad 
D^{(-1)} = B E^{(-1)} B^{-1}
\label{0.9}
\er
where $E^{(-1)}$ is given by (\ref{0.3}) for $n=-1$.  According to $Q$ in (\ref{0.2}) the zero grade subalgebra is generated by
 $ \lie_0 =\{ M_2^{(0)}, \hat c\}$, i.e. 
 \br
B = \exp {(\phi M_2^{(0)} + \nu \hat c)}
\label{0.10}
\er
The time evolution   for $t_{-1}$ is obtained from (\ref{0.8}) and  coincides with 
 the Leznov-Saveliev's equation  \cite{ls} when we identify $(x,\; t_{-1})$ 
 with the light cone coordinates, 
\br
\pa_{t_{-1}} \pa_x BB^{-1} &=&-[E, B E^{(-1)} B^{-1}] -[\bar \psi  G_1^{(1/2)}, \psi B G_2^{(-1/2)}B^{-1}], \nonu \\
\pa_{t_{-1}} \bar \psi  G_1^{(1/2)}&=& [E, \psi B G_2^{(-1/2)}B^{-1}]
\label{0.11}
\er
leading in components to the $N=1$ super sinh-Gordon equations of motion, 
\br
\pa_{t_{-1}} \pa_x \phi &=& 2 \sinh 2\phi + 2 \bar \psi \psi \sinh \phi, \nonu \\
\pa_{t_{-1}} \pa_x \nu &=& \psi \bar \psi  e^{\phi} + (1 - e^{2\phi}), \nonu \\
\pa_{t_{-1}} \bar \psi &=& 2 \psi \cosh \phi, \nonu \\
\pa_x \psi &=& 2 \bar \psi \cosh \phi
\label{0.12}
\er
The above equations are invariant under supersymmetry transformation
\br
\bar \psi^{\pr} = \bar \psi  + \eps \pa_x \phi, \quad \quad  \phi^{\pr} = \phi + \eps \bar \psi
\label{0.13}
\er
The zero group element $B$ defined in  (\ref{0.9})  when parametrized as in (\ref{0.10}) 
establishes a correspondence between relativistic (sinh-Gordon) and non relativistic  (mKdV) field variables, i.e.
\br
u = -\pa_x \phi, \quad \quad \eta = -\pa_x \nu
\label{var}
\er

%%%%%%%%%%%%%%%%%%%%%%%%%%%%%%%%%%%%%%%%%%%%%%%%%%%%%%%%%%%%%%%%%%%%%%%%%%%%%%%%%%%%%%%%%%%%%%%%%%%%%%%%%%%%%%%%%%%%%%%
\section{Dressing and Soliton Solutions}
We now construct the soliton solution for both, mKdV and sinh-Gordon models
from the dressing transformation generated by $\Theta_{\pm}$ 
which relates two solutions of the equations of motion 
written in the zero curvature representation. 
In particular, it
relates  the vacuum and the 1-soliton solutions by a gauge transformation,
\br
 {\cal{A}}_{\mu}= \Theta_{\pm} {\cal{A}}_{\mu}^{ vac}\Theta_{\pm}^{-1} - \( \pa_{\mu}\Theta_{\pm}\) \Theta_{\pm}^{-1}
 \label{dress}
 \er
where 
\br
\Theta_- =  e^{ m(-{1/2})+ m(-1) \cdots }   \quad \quad \Theta_+ = Be^{v({1/2})+ v(1) \cdots }
\label{asoliton}
\er
where $m(-i) \in \lie_{-i}$ and  $v(i) \in \lie_{i}$.
The zero curvature representation implies for pure gauge solutions: 
\br
{\cal{A}}_{\mu}^{ vac }= -\pa_{\mu}T_0 T^{-1}_0 , \quad \quad {\cal{A}}_{\mu}= -\pa_{\mu}T T^{-1}
\label{t0}
\er
which leads to the following relation 
\br
\Theta_-^{-1} \Theta_+  = T_0 gT_0^{-1} \, ,
\label{4.3}
\er
where $g \in \hat {G}$ is an arbitrary constant element of the corresponding affine group.  Suppose 
$T_0 $ represents  the vacuum solution, 
\br
T_0 = \exp (-t_n E^{(n)}) \exp (- x E^{(1)}) , \qquad 
\label{4.4}
\er
i.e.,
\br 
{\cal{A}}_{t_n}^{  vac }= E^{(n)}, \quad \quad  {\cal{A}}_{x}^{  vac }= E^{(1)} 
\label{avac}
\er 
As consequence of (\ref{dress}) with (\ref{avac}) and (\ref{asoliton}) we 
can determine $\Theta_{\pm}$. 
Consider for instance eqn. (\ref{dress}) for ${\cal A}_x$ and $\Theta_-$.   
Its zero  and half grade components  determine the  ${\cal M}$ components of   $m(-1/2)$ and 
$m(-1)$   through  
\br
A_{1/2} = [m(-1/2), E], \quad \quad
A_0 = [m(-1), E] + {1\o 2} [m(-1/2),A_{1/2}]
\label{e}
\er
The  same equation (\ref{dress}), for grades $-1/2$ and $-1$ yields respectively
\br
\pa_x m(-1/2) &=& [m(-3/2), E] + {1\o {2}} \lb m(-1/2),\lb m(-1), E\rb \rb \nonu \\
&+& {1\o {2}}[m(-1), A_{1/2}] + {1\o {3!}} \lb  m(-1/2),\lb  m(-1/2), A_{1/2}\rb \rb \nonu \\
\pa_x m(-1) &=& -{1\o {2}} m(-1/2)\pa_x m(-1/2) + {1\o {2}} \pa_x m(-1/2)m(-1/2) + \lb m(-2), E\rb \nonu \\
&+& {1\o {2}} \lb  m(-1/2),\lb  m(-3/2), E \rb \rb  + {1\o {2}} \lb  m(-1),\lb  m(-1), E \rb \rb + 
{1\o {2}}[m(-3/2), A_{1/2}] \nonu \\
&+& {1\o {3!}} \lb  m(-1/2),\lb  m(-1/2), \lb m(-1), E\rb \rb \rb +
{1\o {3!}} \lb  m(-1/2),\lb  m(-1), A_{1\/2} \rb \rb \nonu \\
&+& {1\o {3!}} \lb  m(-1),\lb  m(-1/2), A_{1/2} \rb \rb + 
{1\o {4!}} \lb  m(-1/2),\lb  m(-1/2), \lb  m(-1/2),A_{1/2} \rb \rb \rb \nonu \\
\label{f}
\er
and determines the kernel, ${\cal K}$, components of   $m(-1/2)$ and 
$m(-1)$   (which, in principle is non local) together with the  image, ${\cal M}$, components of   $m(-3/2)$ and 
$m(-2)$. From $A_0$ and $ A_{1/2}$  given by (\ref{0.5})
we  find   for the super mKdV  
\br 
m(-1/2) &=& \a_1 G_2^{(-1/2)} +\a_2 F_1^{(-1/2)}, \nonu \\
m(-1) &=&  \b_1 M_1^{(-1)} + \b_2 K_1^{(-1)}+ \b_3 K_2^{(-1)}
\label{m}
\er
where  
\br
\a_1 &=& -{1\o 2} \bar \psi, \quad  \quad \a_2  =- {1\o 2} \chi, \quad  \quad \b_1  = {1\o 2} ( u - {1\o 2} \bar \psi \chi ) \nonu \\
\b_2 &=& 
{1\o 4} \int (\bar \psi \pa_x \bar \psi - \chi \pa_x \chi ) dx - {1\o 2} \int
u^2 dx  \quad  \quad 
\b_3 = -{1\o 4} \int ( \bar \psi \pa_x \bar \psi + \chi \pa_x \chi ) dx, \nonu \\
\label{beta}
\er
and $   \pa_x \chi = u \bar \psi $.  
It also leads to the eqn. for $\eta$, 
\br
2\pa_x \eta = u^2 -\pa_x u - \bar \psi \pa_x \bar \psi.
\er 
The full dressing transformation $\Theta_{\pm}$ is then determined  by  considering higher grade terms  of 
(\ref{dress}) with (\ref{avac}) and (\ref{asoliton}).

%%%%%%%%%%%%%%%%%%%%%%%%%%%%%%%%%%%%%%%%%%%%%%%%%%%%%%%%%%%%%%%%%%%%%%%%%%%%%%%%%%%
From eqn. (\ref{4.3})  the following $\tau$-functions are  found,
\br
\tau_0 &=& e^{\nu} = <\l_0|T_0 g T_0^{-1}|\l_0>, \nonu \\
\tau_1 &=& e^{\phi+\nu} = <\l_1|T_0 g T_0^{-1}|\l_1>, \nonu \\
\tau_2 &=& {1\o 2}(\bar \psi - \chi)e^{\nu} = <\l_0|G_1^{(1/2)}T_0 g T_0^{-1}|\l_0>, \nonu \\
\tau_3 &=& {1\o 2}(\bar \psi + \chi)e^{\phi+\nu} = <\l_1|G_1^{(1/2)}T_0 g T_0^{-1}|\l_1>
\label{d.2}
\er
where  $\l_i, \; i=0,1$ denote the first two fundamental weights of $\hat {sl}(2,1)$ satisfying 
\br
\hat c | \l_i> = | \l_i>, \quad \quad M_2^{(0)} | \l_i>  =  \d_{i,1}| \l_i>
\er
and are annihilated by the positive grade generators.
The soliton solution is therefore given  in terms of representations of the $\hat {sl}(2,1)$ affine Lie super algebra, 
\br
\phi = ln \({{\tau_1}\o {\tau_0}}\),  \quad 
\bar \psi = {{\tau_3}\o {\tau_1}} +{{ \tau_2}\o {\tau_0}}, 
\label{d.3}
\er
For the  relativistic sinh-Gordon $t_n = t_{-1} $ whilst for the non relativistic 
mKdV model  $t_n = t_{3}$, $u =-\pa_x \phi$ and  $\eta = -\pa_x \nu$.

The soliton solutions are classified  in terms of the constant element $g$ in (\ref{4.3}) which   
is constructed in   terms of eigenvectors of $E^{(n)}$, i.e.,
\br
[ E^{(2n+1)}, F_{\pm} (\g) ] = \pm 2 \g^{2n+1} F_{\pm}(\g)
\label{d.4}
\er
where 
\br
F_{-}(\g) &=& \sum_{n \in Z} M_1^{(2n+1)}\g^{-2n-1} + (M_2^{(2n)} - {1\o 2}\hat c \d_{n,0})\g^{-2n}, \nonu \\
F_{+}(\g) &=& \sum_{n \in Z} G_1^{(2n+1/2)}\g^{-2n} +G_2^{(2n+3/2)}\g^{-2n-1}
\label{d.5}
\er
%%%%%%%%%%%%%%%%%%%%%%%%%%%%%%%%%%%%%%%%%%%%%%%%%%%%%%%%%%%%%%%%%%%%%%%%%%%%%%%%%%%%%%%%%%%%%%%%%%%%%%%%%%%%%
\subsection{Two Vertex Solution}
Consider  as an illustration the case where 
\br
g = e^{b_1 F_{-}(\g_1)} e^{c_1 F_{+}(\g_3)}
\label{d.6a}
\er
where $b_1$ and $c_1 $ are bosonic and fermionic coefficients respectively.
By virtue of (\ref{d.4}) the explicit space-time dependence on the r.h.s. of (\ref{d.2}) is  
\br
T_0 g T_0^{-1} = 
e^{b_1 \rho^{+}_1(\g_1) F_{-}(\g_1)}
e^{c_1 \rho^{-}_3(\g_3) F_{+}(\g_3)}
\label{d.7a}
\er
where
\br
{\rho}_{i}^{\pm} =e^{\pm(2\gamma_{i}x+2\gamma_{i}^{2n+1} t_{2n+1})}.
\label{d.8a}
\er
The $\tau $ functions (\ref{d.2})  become 
\begin{eqnarray}
\tau_{0}=e^{\nu}
&=&1-\frac{1}{2}b_{1}{\rho}_{1}^{+} + b_1c_1 {\rho}_{1}^{+}{\rho}_{3}^{-}<\l_0 |F_{-}(\g_1) F_{+}(\g_3) |\l_0 >, \nonu \\
\tau_{1}=e^{\phi+\nu}
&=&1+\frac{1}{2}b_{1}{\rho}_{1}^{+}+ b_1c_1 {\rho}_{1}^{+}{\rho}_{3}^{-}<\l_1 |F_{-}(\g_1) F_{+}(\g_3) |\l_1 >, \nonu \\
\tau_{2}=\frac{1}{2}(\bar{\psi}-\chi)e^{\nu} 
&=&c_{1}{\rho}_{3}^{-}\gamma_{3} 
+b_1c_1 {\rho}_{1}^{+}{\rho}_{3}^{-}<\l_0 |G_1^{(1/2)}F_{-}(\g_1) F_{+}(\g_3) |\l_0 >, \nonu \\
\tau_{3}=\frac{1}{2}(\bar{\psi}+\chi)e^{\phi+\nu}
&=&c_{1}{\rho}_{3}^{-}\gamma_{3}
+b_1c_1 {\rho}_{1}^{+}{\rho}_{3}^{-}<\l_1 |G_1^{(1/2)}F_{-}(\g_1) F_{+}(\g_3) |\l_1 >, \nonu \\
\end{eqnarray}
where the matrix elements  can be evaluated from the representation theory of the affine 
$\hat {sl}(2,1)$ super algebra (\ref{d.2}) yielding
\br
<\l_i |F_{-}(\g_1) F_{+}(\g_3) |\l_i > &=& 0 \nonu \\
<\l_i |G_1^{(1/2)} F_{-}(\g_1) F_{+}(\g_3) |\l_i > &=&  {{\g_3}\o {2}} {{(\g_1 +\g_3)}\o {(\g_1 -\g_3)}}(1-2\d_{i,1}), \quad i=0,1
\er
Since, in our formulation, 
the super mKdV belongs to the same hierarchy as the super sinh-Gordon, its  
solution is determined using the same vertex functions (\ref{d.5})  and 
substituting  $\g_i^{-1}t_{-1} $ by $\g_i^3 t_3$  
together with the change of variables $u = -\pa_x \phi,  \eta = - \pa_x \nu$. The 
explicit space-time dependence according to the sinh-Gordon or mKdV are given respectively by,
\br
{\rho}_{i\; S-G}^{\pm}=e^{\pm(2\gamma_{i}x+2\gamma_{i}^{-1}t_{-1})}\quad \quad 
{\rho}_{i\; mKdV}^{\pm}=e^{\pm(2\gamma_{i}x+2\gamma_{i}^{3}t_{3})}
\label{rho}
\er
The corresponding two-vertex solution  for the super mKdV is then given by
\br 
u &=&-\pa_x \phi = -b_1\g_1 {\rho}_{1}^{+}\( {{1}\o {1+{1\o 2} b_1 {\rho}_{1}^{+}}} + {{1}\o {1-{1\o 2} b_1 {\rho}_{1}^{+}}}\),\nonu \\
\bar \psi &=& {{c_1 {\rho}_{3}^{-}\g_3 - b_1c_1 {\rho}_{1}^{+}{\rho}_{3}^{-}\s_{1,3}}\o {1+{1\o 2} b_1 {\rho}_{1}^{+}}}
+{{c_1 {\rho}_{3}^{-}\g_3 + b_1c_1 {\rho}_{1}^{+}{\rho}_{3}^{-}\s_{1,3}}\o {1-{1\o 2} b_1 {\rho}_{1}^{+}}},\nonu \\
\eta &=& -\pa_x \nu ={{b_1 \g_1 {\rho}_{1}^{+}}\o {1-{1\o 2} b_1 {\rho}_{1}^{+}}}, 
\quad \quad \s_{1,3}(\g_1, \g_3) = {{\g_3}\o {2}} {{(\g_1 +\g_3)}\o {(\g_1 -\g_3)}}
\label{sol2ver}
\er
For the particular case where $\g_1 =- \g_3 = k$, $b_1 = -2$ and $c_1 = -{{\xi }\o {k}}$ our solution for $u$ and $\bar \psi$ (\ref{sol2ver})
coincide (after scaling $t_3$) with the one obtained  in \cite{liu}  by  
extending the bilinear approach to the supersymmetric mKdV equation. 
%%%%%%%%%%%%%%%%%%%%%%%%%%%%%%%%%%%%%%%%%%%%%%%%%%%%%%%%%%%%%%%%%%%%%%%%%%%%%%%%%%%%%%%%%%%%%%%%%%%%%%%%%%%%%
\subsection{Four Vertex Solution}
 We now  explicit display  the general 4-vertex solution where 
\br
g = e^{b_1 F_{-}(\g_1)} e^{b_2 F_{-}(\g_2)}e^{c_1 F_{+}(\g_3)}e^{c_2 F_{+}(\g_4)}
\label{d.6}
\er
where $b_i$ and $c_i, \; i=1,2$ are bosonic and fermionic coefficients respectively and
\br
T_0 g T_0^{-1} = 
e^{b_1 \rho^{+}_1(\g_1) F_{-}(\g_1)}e^{b_2 \rho^{+}_2(\g_2) F_{-}(\g_2)}
e^{c_1 \rho^{-}_3(\g_3) F_{+}(\g_3)}e^{c_2 \rho^{-}_4(\g_4) F_{+}(\g_4)}
\label{d.7}
\er
%%%%%%%%%%%%%%%%%%%%%%%%%%%%%%%%%%%%%%%%%%%%%%%%%%%%%%%%%%%%%%%%%%%%%%%%%%%%%%%%%%%%%%%%%%%%%%%%%%%%%%%%%%%%%
 The $\tau$ functions in (\ref{d.2}) become
\begin{eqnarray}
\tau_{0}&=&e^{\nu}
=1-\frac{1}{2}b_{1}{\rho}_{1}^{+}-\frac{1}{2}b_{2}{\rho}_{2}^{+}+
b_{1}b_{2}{\rho}_{1}^{+}{\rho}_{2}^{+}\alpha_{1,2}\nonumber\\
&+&c_{1}c_{2}{\rho}_{3}^{-}{\rho}_{4}^{-}\left(\beta_{3,4}-b_{1}{\rho}_{1}^{+}\delta_{1,3,4}-
b_{2}{\rho}_{2}^{+}\delta_{2,3,4}+b_{1}b_{2}{\rho}_{1}^{+}{\rho}_{2}^{+}\theta_{1,2,3,4}\right),\nonumber\\[12pt] 
\tau_{1}&=&e^{\phi+\nu}
=1+\frac{1}{2}b_{1}{\rho}_{1}^{+}+\frac{1}{2}b_{2}{\rho}_{2}^{+}+
b_{1}b_{2}{\rho}_{1}^{+}{\rho}_{2}^{+}\alpha_{1,2}\nonumber\\
&+&c_{1}c_{2}{\rho}_{3}^{-}{\rho}_{4}^{-}\left(\beta_{3,4}+
b_{1}{\rho}_{1}^{+}\delta_{1,3,4}+b_{2}{\rho}_{2}^{+}\delta_{2,3,4}+
b_{1}b_{2}{\rho}_{1}^{+}{\rho}_{2}^{+}\theta_{1,2,3,4}\right),\nonumber\\[12pt]
\tau_{2}&=&\frac{1}{2}(\bar{\psi}-\chi)e^{\nu}
=c_{1}{\rho}_{3}^{-}\left(\gamma_{3}+b_{1}{\rho}_{1}^{+}\sigma_{1,3}+b_{2}{\rho}_{2}^{+}\sigma_{2,3}+
b_{1}b_{2}{\rho}_{1}^{+}{\rho}_{2}^{+}\lambda_{1,2,3}\right)\nonumber\\
&+&c_{2}{\rho}_{4}^{-}\left(\gamma_{4}+b_{1}{\rho}_{1}^{+}\sigma_{1,4}+
b_{2}{\rho}_{2}^{+}\sigma_{2,4}+b_{1}b_{2}{\rho}_{1}^{+}{\rho}_{2}^{+}\lambda_{1,2,4}\right),\nonumber\\[12pt]
\tau_{3}&=&\frac{1}{2}(\bar{\psi}+\chi)e^{\phi+\nu}
=c_{1}{\rho}_{3}^{-}\left(\gamma_{3}-b_{1}{\rho}_{1}^{+}\sigma_{1,3}-
b_{2}{\rho}_{2}^{+}\sigma_{2,3}+b_{1}b_{2}{\rho}_{1}^{+}{\rho}_{2}^{+}\lambda_{1,2,3}\right)\nonumber\\
&+&c_{2}{\rho}_{4}^{-}\left(\gamma_{4}-b_{1}{\rho}_{1}^{+}\sigma_{1,4}-
b_{2}{\rho}_{2}^{+}\sigma_{2,4}+b_{1}b_{2}{\rho}_{1}^{+}{\rho}_{2}^{+}\lambda_{1,2,4}\right).
\end{eqnarray}
where the coefficients are given by \footnote{we have used the Mathematica program of ref. \cite{laf1}}  
\begin{eqnarray}
\alpha_{1,2}&=&\frac{1}{4}\frac{(\gamma_{1}-\gamma_{2})^{2}}{(\gamma_{1}+\gamma_{2})^{2}},\nonumber\\[12pt]
\beta_{3,4}&=&\gamma_{3}\gamma_{4}\frac{(\gamma_{3}-\gamma_{4})}{(\gamma_{3}+\gamma_{4})^{2}},\nonumber\\[12pt]
\delta_{j,3,4}&=&\frac{\gamma_{3}\gamma_{4}}{2}\frac{(\gamma_{3}-\gamma_{4})}{(\gamma_{3}+
\gamma_{4})^{2}}\frac{(\gamma_{j}+\gamma_{3})}{(\gamma_{j}-\gamma_{3})}\frac{(\gamma_{j}+
\gamma_{4})}{(\gamma_{j}-\gamma_{4})} \qquad (j=1,2),\nonumber\\[12pt]
\sigma_{j,k}&=&\frac{\g_k}{2}\frac{(\gamma_{j}+\gamma_{k})}{(\gamma_{j}-
\gamma_{k})} \qquad (j=1,2) \qquad (k=3,4),\nonumber\\[12pt]
\lambda_{1,2,j}&=&\frac{\gamma_{j}}{4}\frac{(\gamma_{1}-\gamma_{2})^{2}}{(\gamma_{1}+
\gamma_{2})^{2}}\frac{(\gamma_{1}+\gamma_{j})}{(\gamma_{1}-
\gamma_{j})}\frac{(\gamma_{2}+\gamma_{j})}{(\gamma_{2}-\gamma_{j})}, \qquad (j=3,4),\nonumber\\[12pt]
\theta_{1,2,3,4}&=&\frac{\gamma_{3}\gamma_{4}}{4}\frac{(\gamma_{1}-\gamma_{2})^{2}}{(\gamma_{1}+
\gamma_{2})^{2}}\frac{(\gamma_{1}+\gamma_{3})}{(\gamma_{1}-\gamma_{3})}\frac{(\gamma_{2}+\gamma_{3})}{(\gamma_{2}
-\gamma_{3})}\nonumber\frac{(\gamma_{3}-\gamma_{4})}{(\gamma_{3}+\gamma_{4})^{2}}\frac{(\gamma_{1}+
\gamma_{4})}{(\gamma_{1}-\gamma_{4})}\frac{(\gamma_{2}+\gamma_{4})}{(\gamma_{2}-\gamma_{4})}.\nonumber\\
\end{eqnarray}
The solution for the super sinh-Gordon is then given as
\begin{eqnarray}
\phi&=&\ln\left(\frac{1+\frac{1}{2}b_{1}{\rho}_{1}^{+}+\frac{1}{2}b_{2}{\rho}_{2}^{+}+
b_{1}b_{2}{\rho}_{1}^{+}{\rho}_{2}^{+}\alpha_{1,2}}{1-\frac{1}{2}b_{1}{\rho}_{1}^{+}-
\frac{1}{2}b_{2}{\rho}_{2}^{+}+b_{1}b_{2}{\rho}_{1}^{+}{\rho}_{2}^{+}\alpha_{1,2}}\right)\nonumber\\
&+&c_{1}c_{2}\frac{{\rho}_{3}^{-}{\rho}_{4}^{-}\left(\beta_{3,4}+
b_{1}{\rho}_{1}^{+}\delta_{1,3,4}+b_{2}{\rho}_{2}^{+}\delta_{2,3,4}+
b_{1}b_{2}{\rho}_{1}^{+}{\rho}_{2}^{+}\theta_{1,2,3,4}\right)}{1+\frac{1}{2}b_{1}{\rho}_{1}^{+}+
\frac{1}{2}b_{2}{\rho}_{2}^{+}+b_{1}b_{2}{\rho}_{1}^{+}{\rho}_{2}^{+}\alpha_{1,2}}\nonumber\\
&-&c_{1}c_{2}\frac{{\rho}_{3}^{-}{\rho}_{4}^{-}\left(\beta_{3,4}-b_{1}{\rho}_{1}^{+}\delta_{1,3,4}-
b_{2}{\rho}_{2}^{+}\delta_{2,3,4}+b_{1}b_{2}{\rho}_{1}^{+}{\rho}_{2}^{+}\theta_{1,2,3,4}\right)}{1
-\frac{1}{2}b_{1}{\rho}_{1}^{+}-\frac{1}{2}b_{2}{\rho}_{2}^{+}+b_{1}b_{2}{\rho}_{1}^{+}{\rho}_{2}^{+}\alpha_{1,2}},
\label{phi} 
\er
\br
\bar{\psi}&=&c_{1}\frac{{\rho}_{3}^{-}\left(\gamma_{3}-b_{1}{\rho}_{1}^{+}\sigma_{1,3}-
b_{2}{\rho}_{2}^{+}\sigma_{2,3}+
b_{1}b_{2}{\rho}_{1}^{+}{\rho}_{2}^{+}\lambda_{1,2,3}\right)}{1+\frac{1}{2}b_{1}{\rho}_{1}^{+}+\frac{1}{2}b_{2}{\rho}_{2}^{+}
+b_{1}b_{2}{\rho}_{1}^{+}{\rho}_{2}^{+}\alpha_{1,2}}\nonumber\\
&+&c_{2}\frac{{\rho}_{4}^{-}\left(\gamma_{4}-b_{1}{\rho}_{1}^{+}\sigma_{1,4}-b_{2}{\rho}_{2}^{+}\sigma_{2,4}+
b_{1}b_{2}{\rho}_{1}^{+}{\rho}_{2}^{+}\lambda_{1,2,4}\right)}{1+\frac{1}{2}b_{1}{\rho}_{1}^{+}+
\frac{1}{2}b_{2}{\rho}_{2}^{+}+b_{1}b_{2}{\rho}_{1}^{+}{\rho}_{2}^{+}\alpha_{1,2}}\nonumber\\
&+&c_{1}\frac{{\rho}_{3}^{-}\left(\gamma_{3}+b_{1}{\rho}_{1}^{+}\sigma_{1,3}+
b_{2}{\rho}_{2}^{+}\sigma_{2,3}+b_{1}b_{2}{\rho}_{1}^{+}{\rho}_{2}^{+}\lambda_{1,2,3}\right)}{1-\frac{1}{2}b_{1}{\rho}_{1}^{+}
-\frac{1}{2}b_{2}{\rho}_{2}^{+}+b_{1}b_{2}{\rho}_{1}^{+}{\rho}_{2}^{+}\alpha_{1,2}}\nonumber\\
&+&c_{2}\frac{{\rho}_{4}^{-}\left(\gamma_{4}+b_{1}{\rho}_{1}^{+}\sigma_{1,4}+
b_{2}{\rho}_{2}^{+}\sigma_{2,4}+b_{1}b_{2}{\rho}_{1}^{+}{\rho}_{2}^{+}\lambda_{1,2,4}\right)}{1-\frac{1}{2}b_{1}{\rho}_{1}^{+}
-\frac{1}{2}b_{2}{\rho}_{2}^{+}+b_{1}b_{2}{\rho}_{1}^{+}{\rho}_{2}^{+}\alpha_{1,2}},
\label{psi}
\er
\br
\nu&=&\ln\left(1-\frac{1}{2}b_{1}{\rho}_{1}^{+}-\frac{1}{2}b_{2}{\rho}_{2}^{+}
+b_{1}b_{2}{\rho}_{1}^{+}{\rho}_{2}^{+}\alpha_{1,2}\right)\nonumber\\
&+&c_{1}c_{2}\frac{{\rho}_{3}^{-}{\rho}_{4}^{-}\left(\beta_{3,4}-b_{1}{\rho}_{1}^{+}\delta_{1,3,4}
-b_{2}{\rho}_{2}^{+}\delta_{2,3,4}+b_{1}b_{2}{\rho}_{1}^{+}{\rho}_{2}^{+}\theta_{1,2,3,4}\right)}{1-\frac{1}{2}b_{1}{\rho}_{1}^{+}
-\frac{1}{2}b_{2}{\rho}_{2}^{+}+b_{1}b_{2}{\rho}_{1}^{+}{\rho}_{2}^{+}\alpha_{1,2}}.
\end{eqnarray}
where ${\rho}_{i}^{\pm}={\rho}_{i\; S-G}^{\pm}$.

The soliton solutions for the super mKdV are obtained by replacing  ${\rho}_{i}^{\pm}={\rho}_{i\; mKdV}^{\pm}$ 
in (\ref{rho}) and  writing $u=-\pa_x \phi$ and $\eta = - \pa_x \nu$.  
We have verified that our solution for the four-vertex super 
mKdV  agrees with the one found in ref. \cite{liu} when $\g_1=-\g_3 = k_1$, $\g_2=-\g_4 = k_2$, $b_1=b_2 =-2$ and 
$c_i = - {{\xi_i}\o {k_i}}, \;\; i=1,2$ (after scaling $t_3$).

It becomes  clear that the soliton solutions  are classified in terms of the number and in terms of the type of vertices emploied in
constructing  $g$.
Other integrable equations within the same hierarchy and associated to higher grade time evolution, $t_{2n+1}$,
can be   constructed from the zero curvature condition (\ref{0.6}) by replacing  $E^{(1)}$  by  $E^{(2n+1)}$ (given in (\ref{0.3})). 
They all share the same  soliton solutions  for fields 
$u(x, t_{2n+1})=-\pa_x \phi,\;\;  \eta(x, t_{2n+1}) = - \pa_x \nu, \;\; \bar \psi (x, t_{2n+1}) $ with ${\rho}_{i}^{\pm}$ given by
${\rho}_{i}^{\pm}=e^{\pm(2\gamma_{i}x+2\gamma_{i}^{2n+1} t_{2n+1})}$.

%%%%%%%%%%%%
\section*{\sf Acknowledgments}
We are gratefull to H. Aratyn for discussions.
LHY acknowledges support from Fapesp,
JFG and AHZ thank CNPq for a partial support.

%%%%%%%%%%%%%%%%%%%%%%%%%%%%%%%%%%%%%%%%%%%%%%%%%%%%%%%%%%%%%%%%%%%%%%%%%%%%%%%%%%%%%%%%%%%%%%%%%%%%%%%%%%%%%%%%%%%%%%%
\section{Appendix A}

We now discuss how to implement  the relevant {\it subalgebra} of the affine $\hat {sl}(2,1)$ Kac-Moody algebra
in order to construct the integrable hierarchy we are interested in.
Consider the ${sl}(2,1)$ Lie algebra  with generators 
\br
\{ h_1 = {{2\a_1 \cdot H}\o {\a_1^2}}, \; \l_2\cdot H,  \; E_{\pm \a_1}, \;  E_{\pm \a_2},  \; E_{\pm (\a_1+ \a_2)}\} 
\label{a.1}
\er
where $\a_1$,$\a_2$ are the   bosonic   and fermionic simple roots respectively and $\l_2$ is the second fundamental weight. 
The affine $\hat {sl}(2,1)$ structure is implemented by extending each generator $T_a \in  {sl}(2,1)$ to $T_a^{(q)}$ where 
$[ d, T_a^{(q)}] = q  T_a^{(q)}$.
 The relevant {\it subalgebra} of the affine $\hat {sl}(2,1)$ is constructed as follows.
The grade one constant element $E^{(1)}$ be given in (\ref{0.3}) decomposes the affine algebra into
\br
{\cal K}_{Bose} &=& \{ K_1^{(2n+1)} = -(E_{a_1}^{(n)} +E_{-a_1}^{(n+1)} ), \quad K_2^{(2n+1)} = \l_2 \cdot H^{(n+1/2)})\}\nonu \\
{\cal M}_{Bose} &=&\{M_1^{(2n+1)} = -E_{\a_1}^{(n)} +E_{-\a_1}^{(n+1)}, \quad M_2^{(2n)} =h_1^{(n)} \}
\label{a.2}
\er
and the fermionic sector 
\br
{\cal K}_{Fermi} =  F_1^{(2n+3/2)} &=& (E_{\a_1+\a_2}^{(n+1/2)} -E_{\a_2}^{(n+1)} )+ 
(E_{-\a_1-\a_2}^{(n+1)} -E_{-\a_2}^{(n+1/2)} ) \nonu \\
 F_2^{(2n+1/2)} &=& -(E_{\a_1+\a_2}^{(n)} -E_{\a_2}^{(n+1/2)} )+ 
(E_{-\a_1-\a_2}^{(n+1/2)} -E_{-\a_2}^{(n)} )  \nonu \\
{\cal M}_{Fermi} =G_1^{(2n+1/2)} &=& (E_{\a_1+\a_2}^{(n)} +E_{\a_2}^{(n+1/2)} )+ 
 (E_{-\a_1-\a_2}^{(n+1/2)} +E_{-\a_2}^{(n)} ) \nonu \\
G_2^{(2n+3/2)} &=& -(E_{\a_1+\a_2}^{(n+1/2)} +E_{\a_2}^{(n+1)} )+ 
(E_{-\a_1-\a_2}^{(n+1)} +E_{-\a_2}^{(n+1/2)} ) 
\label{a.3}
\er
The algebra is then given by the commutators
\br
& \lb K_1^{(2m+1)}, K_1^{(2n+1)} \rb = \hat c (m-n)\d_{m+n+1,0}, & \quad \quad 
\lb K_1^{(2m+1)}, K_2^{(2n+1)} \rb = 0, \nonu \\
& \lb K_1^{(2m+1)}, M_1^{(2n+1)} \rb = -2M_2^{(m+n+1)} -\hat c (m+n)\d_{m+n+1,0}, & \quad \quad 
\lb K_2^{(2m+1)}, M_1^{(2n+1)} \rb = 0, \nonu \\
& \lb M_2^{(2m)}, K_1^{(2n+1)} \rb = 2M_1^{(2m+2n+1)}, & \quad \quad 
\lb M_2^{(2m)}, K_2^{(2n+1)} \rb = 0, \nonu \\
& \lb M_2^{(2m)}, M_2^{(2n)} \rb = \hat c (m-n)\d_{m+n,0}, & \quad \quad 
\lb M_2^{(2m)}, M_1^{(2n+1)} \rb = 2K_1^{(2m+2n+1)} \nonu \\
& \lb M_1^{(2m+1)}, M_1^{(2n+1)} \rb = \hat c (n-m)\d_{m+n+1,0}, & \quad \quad 
\lb K_2^{(2m+1)}, K_2^{(2n+1)} \rb = \hat c (n-m)\d_{m+n+1} \nonu \\
\label{a.4}
\er
and
\br
& \lb K_1^{(2m+1)}, F_1^{(2n+3/2)} \rb = -F_2^{(2(m+n+1)+1/2)}, & \quad \quad 
\lb K_1^{(2m+1)}, G_1^{(2n+1/2)} \rb = G_2^{(2(m+n)+3/2)}, \nonu \\
& \lb K_1^{(2m+1)}, F_2^{(2n+1/2)} \rb = -F_1^{(2(m+n)+3/2)}, & \quad \quad 
\lb K_1^{(2m+1)}, G_2^{(2n+3/2)} \rb = G_1^{(2(m+n+1)+1/2)}, \nonu \\
& \lb K_2^{(2m+1)}, F_1^{(2n+3/2)} \rb = F_2^{(2(m+n+1)+1/2)}, & \quad \quad 
\lb K_2^{(2m+1)}, G_1^{(2n+1/2)} \rb = G_2^{(2(m+n)+3/2)}, \nonu \\
& \lb K_2^{(2m+1)}, F_2^{(2n+1/2)} \rb = F_1^{(2(m+n)+3/2)}, & \quad \quad 
\lb K_2^{(2m+1)}, G_2^{(2n+3/2)} \rb = G_1^{(2(m+n+1)+1/2)}, \nonu \\ 
 & \lb M_1^{(2m+1)}, F_1^{(2n+3/2)} \rb = G_1^{(2(m+n+1)+1/2)}, & \quad \quad 
\lb M_1^{(2m+1)}, G_1^{(2n+1/2)} \rb = -F_1^{(2(m+n)+3/2)}, \nonu \\
& \lb M_1^{(2m+1)}, F_2^{(2n+1/2)} \rb = G_2^{(2(m+n)+3/2)}, & \quad \quad 
\lb M_1^{(2m+1)}, G_2^{(2n+3/2)} \rb = -F_2^{(2(m+n+1)+1/2)}, \nonu \\
& \lb M_2^{(2m)}, F_1^{(2n+3/2)} \rb = -G_2^{(2(m+n)+3/2)}, & \quad \quad 
\lb M_2^{(2m)}, G_1^{(2n+1/2)} \rb = -F_2^{(2(m+n)+1/2)}, \nonu \\
& \lb M_2^{(2m)}, F_2^{(2n+1/2)} \rb = -G_1^{(2(m+n)+1/2)}, & \quad \quad 
\lb M_2^{(2m)}, G_2^{(2n+3/2)} \rb = -F_1^{(2(m+n)+3/2)}, \nonu \\
\label{a.5}
\er 
and the anticommutators
 \br
 \lb F_1^{(2m+3/2)}, F_1^{(2n+3/2)} \rb_{+} &=& 2(K_2^{(2(m+n+1)+1)} +K_1^{(2(m+n+1)+1)} ), \nonu \\ 
\lb F_1^{(2m+3/2)}, F_2^{(2n+1/2)} \rb_{+} &=& \hat c(2m-2n+1)\d_{m+n+1,0}, \nonu \\
 \lb F_1^{(2m+3/2)}, G_1^{(2n+1/2)} \rb_{+} &=& 2M_2^{(2(m+n+1))}+\hat c(2m+2n+1)\d_{m+n+1,0},   \nonu \\
\lb F_1^{(2m+3/2)}, G_2^{(2n+3/2)} \rb _{+}&=& -2M_1^{(2m+2n+3)}, \nonu \\
 \lb F_2^{(2m+1/2)}, F_2^{(2n+1/2)} \rb_{+} &= &-2(K_2^{(2m+2n+1)}+K_1^{(2m+2n+1)}),  \nonu \\ 
\lb F_2^{(2m+1/2)}, G_1^{(2n+1/2)} \rb_{+} &=& 2M_1^{(2m+2n+1)}, \nonu \\
 \lb F_2^{(2m+1/2)}, G_2^{(2n+3/2)} \rb_{+} &=& -2M_2^{(2m+2n+2)}-\hat c (2m+2n+1)\d_{m+n+1,0},  \nonu \\
\lb G_1^{(2m+1/2)}, G_1^{(2n+1/2)} \rb_{+} &=& 2(K_2^{(2m+2n+1)} - K_1^{(2m+2n+1)}), \nonu \\
\lb G_1^{(2m+1/2)}, G_2^{(2n+3/2)} \rb_{+} &=& \hat c (2m-2n-1)\d_{m+n+1,0},  \nonu \\ 
\lb G_2^{(2m+3/2)}, G_2^{(2n+3/2)} \rb_{+} &=& -2(K_2^{(2m+2n+3)} - K_1^{(2m+2n+3)}) \nonu \\
 \label{a.6}
\er
where the index $l$ in $K_i^{(l)}, M_i^{(l)}, F_i^{(l)}$ and $G_i^{(l)}$  denote their grade
 with respect to $Q$ given in (\ref{0.2}).

%%%%%%%%%%%%%%%%%%%%%%%%%%%%%%%%%%%%%%%%%%%%%%%%%%%%%%%%%%%%%%%%%%%%%%%%%%%%%%%%%%%%%%%%%%%%%%%%%%%%%%%%%%%%%

\end{document}